\documentstyle[eqsecnum,array,multicol,prb,aps,psfig]{revtex}
\draft
\tighten
\begin{document}

\title{Vortex mediated microwave absorption in superclean layered superconductors.}
\author{A.~ A.~ Koulakov and A.~I.~Larkin}
\address{Theoretical Physics Institute, University of Minnesota,
Minneapolis, Minnesota 55455}

\date{\today}

\maketitle
\begin{abstract}

In the superclean case the spectrum
of vortex core excitations in the presence of disorder is not random but 
consists of two series of equally-spaced levels.\cite{Larkin97}
The I-V characteristics of 
such superconductors displays many interesting phenomena. A series 
of resonances is predicted at frequencies commensurate with 
the spacing of the vortex excitations. These resonances reveal an 
even-odd anomaly. In the presence of one weak impurity the excitation
levels can approach each other and almost cross. 
Absorption at  low frequencies is identified with the resonances 
arising in this case. The results of such microscopic theory
coincide up to the order of magnitude with both the theory employing 
kinetic equation\cite{KK} and the experiment.\cite{Matsuda94}
The non-linear effects associated with Zener transitions  
in such crossings are studied. 
These phenomena can be used as a probe of vortex core excitations.

\end{abstract}
\begin{multicols}{2}

%
%
%
%

\section{Introduction}
\label{Introduction}

High-$T_c$ superconductors (HTSC) in normal state have many anomalous properties, 
differing them from normal metals. For example, the relaxation time
depends on temperature according to the law $\tau^{-1}\approx T/2$
(assuming $\hbar=k_{\rm B}=1$).
It is quite interesting if these properties remain anomalous below
the transition, when superconductivity hinders studying them.
Vortex cores retain a lot of information about the normal state.
However, even in the BCS model the properties of vortex cores in 
superclean superconductors at low temperatures are studied insufficiently. 
The purpose of this work is to fill this gap.

It is believed that the dissipation in the mixed state of type II superconductors
is associated
with the low lying excitations arising inside the vortex cores, while they
are dragged through the sample by the Lorentz forces~\cite{KK}. 
Studying dissipation can therefore shine some light 
at both the spectrum and the relaxation mechanism of these excitations. 

Although the excitation spectrum 
was calculated long ago\cite{deGennes89}, experimentalists 
approached the possibility of its observation only recently.
\cite{Matsuda94,Harris94}
To this end the excitation level spacing $\omega_0$ has to significantly 
exceed the broadening of levels due to various relaxation mechanisms $1/\tau$.
This superclean case $\omega_0\tau \gg 1$ became possible in
Cu based high-$T_c$ superconductors,
as due to a very short coherence length $\xi$, both $\omega_0\sim 1/m\xi^2$
is large and the amount of disorder in the vortex core,
contributing to the level broadening, is  small.
It should be noted that the measurements in 
Refs.~\onlinecite{Matsuda94,Harris94}
are done at microwave frequencies to avoid the effects associated with
pinning.

The {\em DC} absorption in superclean superconductors has been studied
theoretically in a number of papers.~\cite{Larkin97,KK,Guinea95}
Among them the most notable is Ref.~\onlinecite{KK}, in which
the excitations were treated in the quasiclassically 
employing the kinetic equation in the $\tau$-
approximation. The result of Ref.~\onlinecite{KK} 
for s-wave superconductors is
\begin{equation}
\rho_{\rm xx}=\frac  {B} {n_s\left|e\right|c} \frac {1} {\omega_0 \tau}
\label{sigmaKK}
\end{equation}
It should be noted however that contemporary high-$T_c$ superconductors
are extremely anisotropic. For this reason the excitations in vortices
become quasi two-dimensional (2D). It is known that in 2D the kinetic 
equation describes the scattering on impurities poorly. 
This is because a particle has a large probability to return
to the original position and scatter on the same impurity many times.
Therefore this problem can be treated i.e. considering the transitions
between discrete levels.

Such treatment was realized in Ref.~\onlinecite{Feigelman97}.
They used the random matrix theory to describe the transitions
between excitation levels. Therefore their theory is not applicable
to the superclean case.

The effects associated with the discreteness of the excitation spectrum
in the superclean case have been added to consideration 
in Refs.~\onlinecite{Larkin97,Guinea95}.
According to these references the discreteness of levels
makes the absorption non-ohmic.
Ref.~\onlinecite{Guinea95} considers the shake up created by a 
weak impurity passing through the vortex core using the Fermi golden rule.
They identify a critical velocity {\em below} which the non-ohmic effects should
become pronounced. It can be estimated as 
\begin{equation}
v_c = \frac{\omega_0} {k_{\rm F}},
\end{equation}
where $k_{\rm F}$ is the Fermi momentum.
Indeed the impurity passing through the core at velocity $v$ can create
transitions between states separated by $\Omega = v/\lambda_{\rm F}$,
as $\lambda_{\rm F}$ is the characteristic spacial frequency of the 
wavefunctions of the excitations.
In the case $\Omega \lesssim \omega_0$ no transitions can occur
and the absorption is exponentially small.
The authors of Ref.~\onlinecite{Larkin97} argue that 
even if impurity is weak, as long as 
the Born parameter of the impurity $\theta \gg \lambda_{\rm F}/\xi$, 
due to the special form of the excitation wavefunctions,
the levels inside the vortex always cross. These crossings occur
in the so-called {\em dissipative region}, when impurity is at the distances 
$a \approx \xi \theta$ from the center of the vortex.
Here the Born parameter $\theta \lesssim 1$.
The excitations in this case occur due to Landau-Zener
transitions happening in the level crossings. 
Due to this effect the absorption at $v<v_c$ is not exponentially small
but is much {\em larger} than given by Eq.~(\ref{sigmaKK}).

It should be emphasized that the calculations in 
Refs.~\onlinecite{Larkin97,KK,Guinea95} have been done
for the DC case. The experiments in 
Refs.~\onlinecite{Matsuda94,Harris94} are done at microwave frequencies
($\omega \approx 1.4K$). 
The $AC$ absorption in the superclean case has been studied
in Refs.~\onlinecite{Kopnin78}, \onlinecite{Kopnin97}, 
and \onlinecite{Hsu93}. All these studies are done using the 
kinetic equation in the $\tau$-approximation.
This paper is dedicated to the microscopic study of the 
AC absorption in the superclean case. 
We adopt the mechanism of absorption used in 
Refs.~\onlinecite{Larkin97,Guinea95}, i. e. the
motion of vortex relative to the impurities brings
about transitions of the excitations to the higher energy levels.

We study the global diagram of absorption as a function of frequency
and amplitude of the applied current. 
We find that if the energy relaxation
time $\tau_{\varepsilon}$ is  large, the region on the diagram 
where kinetic equation gives the correct order of magnitude
of the result becomes small.

The outline of our paper is as follows. 
In Sec.~\ref{Excitations} we review some basic facts 
from Ref.~\onlinecite{Larkin97} about the
vortex core excitations in the presence of impurities
in superclean layered superconductors. It is easy to see that
in the superclean case the number of impurities per vortex core
per crystalline layer can be estimated as $N_i \sim 1/\omega_0 \tau \theta^2$.
If $\theta \approx 1$ then it is very improbable to have more
than one impurity inside the core in one layer. Another simplification
comes from the fact that if coupling between layers is small (open Fermi surface),
the excitation spectrum in the presence of impurity can be 
calculated independently for every layer. For this reason
Ref.~\onlinecite{Larkin97} treats the excitations in the presence of impurity
as belonging to one two-dimensional layer.
As a result they obtain that in the presence of impurity
the usually equidistant spectrum of excitations, pertinent to the
two-dimensional clean vortex core, ceases to be equidistant.
However the spectrum remains to be strongly correlated.
It is shown that the system of odd levels and the system of even ones 
separately continue to be equidistant with the level spacing $2\omega_0$ in
each individual subsystem.  

In Sec.~\ref{Resonances} be describe the resonances occurring in
vortices under the influence of low amplitude, high frequency
field. The amplitude of vortex motion $x_0$ is assumed to be much
smaller than $\lambda_{\rm F}$, and frequency of external field $\omega$
is comparable or larger than $\omega_0$. 
We argue that the shape of resonant curves reveals
an even-odd anomaly. If $\omega\approx2n\omega_0$, where $n$ is integer,
the transitions occur only within each individual subsystem
of even or odd levels. In this case the resonance is very sharp,
with the resonant curve determined by the remnant inelastic processes.
If on the other hand $\omega\approx\left(2n+1\right)\omega_0$
the transitions between two subsystems of even and odd levels can occur.
In this case the resonant frequency depends on the position of the impurity,
and after averaging over this position the
resonant curve of absorption becomes smeared. 

In Sec.~\ref{Small_amplitude} we study the small amplitude low frequency
absorption. In this case the transition can occur only in dissipative regions,
where impurity makes even and odd levels cross. The result
for ${\rm Re}\sigma_{xx}$ obtained in this case coincides with
Eq.~(\ref{sigmaKK}) in the order of magnitude.
It is therefore purely ohmic.

The non-linear effects are associated with an increase of the amplitude $x_0$.
They are of two types. The first is attributed to the saturation of
energy absorption at long times~\cite{Wilkinson92}. 
It therefore effectively decreases the magnitude of energy dissipation. 
This non-linear effect can be neglected if $\omega\tau_{\epsilon} \ll 1$,
where $\tau_{\epsilon}$ is the time of energy relaxation. 
In the latter case another non-linear effect becomes important.
It arises due to Landau-Zener transitions between the crossing even 
and odd levels, as discussed in Ref.~\onlinecite{Larkin97}. 
It therefore leads to an {\em increase} of absorption with respect to
Eq.~(\ref{sigmaKK}). In Sec.~\ref{Non-linear} we present a phase diagram
of various regimes of dissipation arising in this case. 

Sec.~\ref{Conclusions} is dedicated to our conclusions. We discuss
the possible corrections to our results brought about by interlevel coupling,
pinning, and d-wave order parameter. 
We compare our results to the existing experiment and discuss conditions
at which resonances and non-linear effects can be observed.

%
%
%
%

\section{Vortex core excitations in the presence of an impurity.}
\label{Excitations}

In this Section we briefly review some facts about excitations inside the
vortex core. They can be described by the Bogolyubov equations\cite{deGennes89}:
\begin{equation}
\hat{\cal H}
\left(
\begin{array}{l}
u \\ \\ v
\end{array}
\right)
= E
\left(
\begin{array}{l}
u \\ \\ v
\end{array}
\right),
\label{Bog_Eq}
\end{equation}
where 
\begin{equation}
\hat{\cal H}=
\left(
\begin{array}{ll}
{\displaystyle 
\frac {\bbox{p}^2}{2m} + V\left( \bbox{r}-\bbox{a} \right) - \mu; 
} 
& {\displaystyle \Delta \left( \bbox{r} \right)} \\ \\
{\displaystyle
\Delta \left( \bbox{r}\right)^*; 
} & {\displaystyle 
- \frac {\bbox{p}^2}{2m} - V\left( \bbox{r} -\bbox{a} \right) + \mu 
}
\end{array}
\right).
\label{Bog_Ham}
\end{equation}
Here $\Delta \left(\bbox{r} \right)$, $\mu$, $V\left(r \right)$, and $\bbox{a}$
are the order parameter, the chemical potential, the impurity potential, and
the position of the impurity respectively. 
As it is mentioned in the introduction, in the superclean case we
can consider no more than one impurity per vortex, per layer. 
We will assume that the magnetic field is weak ($B\ll H_{\rm C2}$)
and therefore can be neglected in Eq.~(\ref{Bog_Ham}). We will also assume 
the s-wave order parameter to have the form
\begin{equation}
\Delta \left(\bbox{r} \right) = \Delta \left(r \right) e^{i\phi},
\end{equation}
where $r$ and $\phi$ are the polar coordinates.

The low energy excitation spectrum without impurity is well 
known:~\cite{deGennes89}
\begin{equation}
E^0_n = -\omega_0 \left( n- \frac 12 \right),
\label{Energy0}
\end{equation}
where
\begin{equation}
\begin{array} {l}
{\displaystyle
\omega_0 = \frac {
\displaystyle \int_0^{\infty} 
\frac{\Delta \left(r \right) dr} {r} e^{-2K\left( r\right)}
} 
{\displaystyle
k_{\rm F}
\int_0^{\infty} 
dr e^{-2K\left( r\right)}
},
} \\ \\
{\displaystyle
K\left( r\right) = \frac 1{v_{\rm F}}\int_0^r dr' \Delta \left(r' \right)}, \\ \\
{\displaystyle
n=0, \pm 1, \pm 2 \ldots}.
\end{array}
\end{equation}

The corresponding wavefunctions are given by
\begin{equation}
\left(
\begin{array}{l}
u_n \\ \\ v_n
\end{array}
\right)=
Ce^{-K\left( r\right)}
\left(
\begin{array}{l}
{\displaystyle
e^{in\phi}J_n\left( k_{\rm F} r\right)
} \\ \\ 
{\displaystyle
-e^{i\left(n-1\right)\phi}J_{n-1}\left( k_{\rm F} r\right)
}
\end{array}
\right),
\label{WaveFunctions}
\end{equation}
with $J_n\left( x\right)$ being the Bessel function
and $C$ being the normalization constant:
\begin{equation}
C^2=\frac{k_{\rm F}}{\displaystyle
4 \int_0^{\infty} e^{ -2K\left(r\right) dr } }.
\end{equation}
If Kramer-Pesch effect takes place\cite{Kramers74} 
at  low temperatures ($T \ll T_c$) 
$\Delta (r) \approx \Delta (r=\infty) \equiv \Delta_{\infty},~r \gg \xi \frac{T}{T_c}$. 
Therefore, $K(r)$ is given by 
\begin{equation}
K(r) = \frac {\Delta_{\infty}}{v_{\rm F}} r,~~r \gg \xi \frac{T}{T_c}.
\label{KramersK}
\end{equation}   
Consequently
\begin{equation}
\begin{array}{l}
{\displaystyle C^2 = \frac{m\Delta_{\infty}}{2}  } \\ \\
{\displaystyle \omega_0=\frac{\Delta_{\infty}^2}{\varepsilon_{\rm F}}
\ln\left( \frac{T_c}{T} \right)}
\end{array}
\label{KramersO}
\end{equation}

The excitation spectrum in the presence of a short range 
impurity at point $\bbox{a}$ has been obtained 
in Ref.~\onlinecite{Larkin97}. The energy spectrum is given by
the following equation
\begin{equation}
\cos\left( \frac {\pi E}{\omega_0} \right)=
-\frac {\displaystyle
4\pi\omega_0I_1
}{\displaystyle
4\omega_0^2+\pi^2\left|I\right|^2
}
\label{Energy}
\end{equation}
where
\begin{equation}
I=I_2+iI_1=\frac {2C^2}{\pi k_{\rm F}a}
e^{-2K\left( a\right)} \tilde{V}\left( 2k_{\rm F}\right)
e^{2ik_{\rm F}a}
\label{Is}
\end{equation}
and $\tilde{V}\left( q \right)$ is the Fourier transform of the 
impurity potential. In the derivation of (\ref{Energy}) it has been
assumed that $a \gg k_{\rm F}^{-1}$.

\begin{figure}
\centerline{
\psfig{file=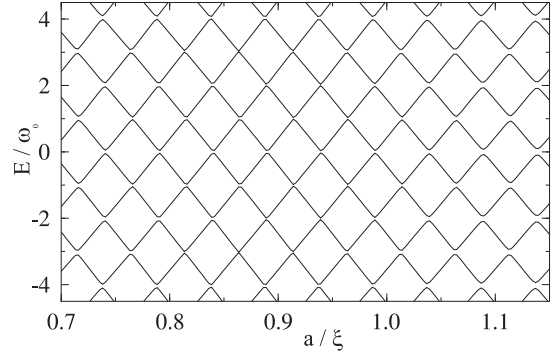,height=1.9in,bbllx=60pt,bblly=103pt,bburx=489pt,bbury=369pt}
}
\vspace{0.1in} 
\setlength{\columnwidth}{3.2in}
\centerline{\caption{The excitation energy levels
as functions of the distance of the impurity from the vortex center
\protect{$a$}. The parameters used are: \protect{$\lambda_{\rm F}/\xi = 0.1$ 
and $\theta\Delta_{\infty}/\omega_0 k_{\rm F} \xi = 0.9$} [see explanation following 
Eq.~(\ref{AnticrossingEnergy}) below]. 
Only the vicinity of the dissipative region is shown.
\label{fig10}
}}
\vspace{-0.1in}
\end{figure}

Note that in the absence of impurity ($I_1 = I = 0$) the 
spectrum given by Eq.~(\ref{Energy}) is equidistant and coincides 
with Eq.~(\ref{Energy0}). However, in the presence of impurity
the spectrum ceases to be equidistant. This is illustrated in Fig.
~\ref{fig10}, where the energy levels are shown as functions of the 
distance of the impurity from the center of the vortex. 

%
%

However, the spectrum remains strongly correlated. It is easy to see from 
Eq.~(\ref{Energy}) and from the Fig.~\ref{fig10} that it comprises two 
series of equidistant levels. The spacing within each series is $2\omega_0$,
while with respect to each other they are shifted by a phase depending
on the impurity. 

Another feature of the spectrum evident from the Figure is that
when the impurity is close to the center it bring about periodic anticrossings
of the levels. The minimum distance between levels in such anticrossings 
$\delta E$, according to Ref.~\onlinecite{Larkin97}, 
has a minimum at point $a=a_0$ and is determined by the equation:
\begin{equation}
\begin{array}{l}
{\displaystyle
\delta E \approx \frac {d\left|I\right|}{da} \left| a - a_0 \right|}, \\ \\
{\displaystyle
\left| I\left( a_0 \right) \right| = \frac {2\omega_0} {\pi} }.
\end{array}
\label{AnticrossingEnergy}
\end{equation}
Eq.~(\ref{AnticrossingEnergy}) together with Eqs.~(\ref{KramersO}) and (\ref{Is}) give
\begin{equation}
a_0 = \frac{\theta \Delta_{\infty}}{k_{\rm F}\omega_0} = \frac {\pi \theta}{2 \ln\left( T_c/T \right)} \xi.
\end{equation}
Here $\theta = m\tilde{V}\left( 2k_{\rm F}\right)$ is the Born parameter.
The region of the vortex near $a=a_0$ is the region where levels approach
each other very closely. For this reason Zener transitions are very probable
there. Therefore it was called in Ref.~\onlinecite{Larkin97}
{\em the dissipative region}.

%
%
%
%

\section{Motion of the vortices}
\label{Motion}

Below we recollect a few facts pertinent to the absorption
by vortices.
We consider the system of vortices in an alternating 
electric field oriented in the plane of the layers.
The magnetic field $\bbox{B}$ is perpendicular to the layers.
If pinning is negligible the velocity of all of the vortices is the same. 
Let us denote it
\begin{equation}
\bbox{v}\left( t\right) =
\bbox{v}\cos \left(\omega t \right).
\end{equation}
Therefore the position of impurity relative to the vortex is given by
\begin{equation}
\bbox{a} = \bbox{\bar{a}} + \bbox{x}_0 \sin \left( \omega t \right),
\end{equation}
where $\bbox{x}_0$ is the amplitude of vibrations related
to the electric field by
\begin{equation}
x_0 = \frac {v}{\omega}.
\end{equation}

It is convenient to write the Schroedinger equation for the time-dependent 
hamiltonian $\hat{\cal H}=\hat{\cal H}\left( t\right)$ 
[see Eq.~(\ref{Bog_Ham})] in the basis of eigenfunctions of this
hamiltonian considering time a parameter.
These eigenfunctions $\left|n\left( t\right)\right>$ and the 
corresponding eigenvalues $E_n\left( t\right)$ can be used to 
obtain the differential equation for the occupation of 
these states\cite{Galitskii63}
\begin{equation}
\dot{c}_n = \sum_{m\neq n} c_m 
\frac{\left< m\right| \frac {\partial \hat{\cal H}}{\partial t} \left| n \right>} 
{E_m\left( t\right) - E_n\left( t\right)}
c_m e^{i\left[ \phi_m\left( t\right) - \phi_n\left( t\right) \right]}.
\label{GalitskiiEq}
\end{equation}
Here $\phi_n = \int_0^t E\left( t'\right)dt'$. 


In Sec.~\ref{Resonances} and \ref{Small_amplitude} 
we will consider cases when 
$x_0 \ll \lambda_{\rm F}$.
The transition probability between levels
$m$ and $n$ in this case can determined by the Fermi golden rule
\begin{equation}
\begin{array}{ll}
{\displaystyle
w_{m \rightarrow n} }&{ \displaystyle = \frac {\pi} 2 
\left| \left< m\right| \delta { \hat{\cal H}} \left| n \right> \right|^2 } \\ \\
{}&{\displaystyle \times \delta \left[ \omega   
- E_n\left(\bar{\bbox{a}}\right) + E_m\left(\bar{\bbox{a}}\right) \right].}
\end{array}
\label{GoldenRule}
\end{equation}
Here the perturbation to the hamiltonian in the system of reference
moving with the condensate
\begin{equation}
\delta {\hat{\cal H}}
=\left(
\begin{array}{ll}
{\displaystyle
\bbox{x}_0 \partial V 
/ {\partial \bbox{r}}; 
}&{\displaystyle
0
} \\ \\
{0; }&{\displaystyle
-\bbox{x}_0 \partial V / {\partial \bbox{r}};}
\end{array}
\right)
\label{HamiltDeriv}
\end{equation}
The energy absorption due to one impurity is given by
\begin{equation}
Q_1 \left(\bar{\bbox{a}}\right) =
\omega \sum_{n, m} w_{m \rightarrow n}
\left[ f_n - f_m \right],
\label{one}
\end{equation}
where $f_n=f\left(E_n\right)=\left[ \exp\left(\frac{E_n-\mu}T\right)+1\right]^{-1}$
is the Fermi distribution function.
The absorption averaged over many impurities and vortices is determined by
\begin{equation}
Q=n_{\rm i} n_{\rm v} \int d^2 \bar{\bbox{a}} Q_1 \left(\bar{\bbox{a}}\right).
\label{many}
\end{equation}
Here $n_{\rm i}$ is the three-dimensional (3D) concentration of impurities,
while $n_{\rm v}$ is the 2D concentration of vortex lines.

Let us now establish the connection between the electric field in the
plane of the layers $\bbox{\cal{E}} (t)$ and the velocity of the vortices
$\bbox{v} (t)$. The AC electric field in a superconductor can be 
written as a sum of two terms:
\begin{equation}
\bbox{\cal{E}} = - \frac 1{4\pi\lambda^2} 
\frac {d\bbox{j}_s}{dt} - \bbox{v\times B}/c,
\label{EquationWithVl}
\end{equation}
where $\lambda$ is the London penetration depth and 
$\bbox{j}_s$ is the supercurrent.
Here the first term is the London electric field 
in the inertial system of reference in which vortices are not moving.
The other term arises due to the flux carried by the vortices with 
velocity $\bbox{v}$.

Let us supplement this formula with the general equations 
for the electric field and dissipation:
\begin{equation}
\begin{array}{l}
{\bbox{\cal{E}} = \bbox{\hat{\rho}j},} \\ \\
{Q=\rm{Re} \rho_{\rm xx} \overline{j^2}},
\end{array}
\label{GeneralEq}
\end{equation}
where the line denotes the time averaging.
Further we consider the case of small temperatures ($T\ll T_c$),
when the relative contribution to the current from normal electrons is 
exponentially small. Therefore we can assume $j\approx j_s$.

In the superclean case $\rm{Im} \rho_{xx}$ and $\rho_{xy}$ are weakly dependent
on the density of impurities. Hence they can be fairly well approximated by the 
values for clean system. Another simplification appears if the spectrum of
electrons is isotropic. Then $\bbox{j}_s=en_s\bbox{v}_s$ and 
$\lambda^{-2} = 4\pi e^2n_s/mc^2$. Eq.~(\ref{GeneralEq}) can then be rewritten 
as the equation of motion of the center of mass of the condensate:
\begin{equation}
{m\dot{\bbox{v}}_s = e\bbox{\cal{E}} 
+ e\bbox{v}_s\times \bbox{B}/c - \eta \bbox{v}_s}. 
\label{ParticularEq}
\end{equation}
Here $\eta$ is the effective viscosity. Calculation of this viscosity
as a function of frequency and velocity is the purpose of this work. Notice
that it has both real and imaginary components. The real  
part can be calculated from the absorption:
\begin{equation}
{Q (\omega) = {\rm Re} \eta \overline{v_s^2}.}
\end{equation}
The imaginary part is somewhat
irrelevant due to the fact that it is compared to the London term
[left-hand side of Eq.~(\ref{ParticularEq})]. 
At the same time ${\rm Im}\eta$ has to vanish in the absence
of impurities as in the opposite case it would shift the position 
of the cyclotron resonance, violating the Kohn's theorem\cite{Kohn61}.
Therefore in the superclean case
it is small compared to the left-hand side of Eq.~(\ref{ParticularEq})
and can be neglected.

Eq.~(\ref{EquationWithVl}) together with Eq.~(\ref{ParticularEq})
result in the following equation for the vortex motion:
\begin{equation}
\bbox{v}_s - \bbox{v} = \bbox{\hat{z}} \times \bbox{v}_s\eta/m\omega_c.
\label{VsMinusVl}
\end{equation}
Here $\omega_c=|e|B/mc$ is the cyclotron frequency.
In the superclean case the ratio $\eta/m\omega_c$ is proportional to the
density of impurities. 
In the majority of cases considered below this quantity is small.
For example the kinetic equation at zero frequency gives 
the following estimate for this quantity:
\begin{equation}
\eta/m\omega_c = 1/\omega_0\tau \ll 1.
\label{Gamma}
\end{equation} 
Our calculations discussed below confirm this result.
Therefore for the superclean case it is plausible to assume that $\bbox{v}$
is  close to $\bbox{v}_s$ [see Eq.~(\ref{VsMinusVl})], i.e. the 
condensate and the vortices almost do not move with respect to each other.

Eq.~(\ref{ParticularEq}) can also be used to obtain an expression for the
resistivity tensor:
\begin{equation}
\begin{array}{l}
{\displaystyle \rho_{\rm xx} = \frac {B}{n_s|e|c} 
\left( -i\frac{\omega}{\omega_c} + \frac {\eta}{m\omega_c}\right),} \\ \\
{\displaystyle \rho_{\rm xy}(\omega) = \frac {B}{n_sec}.}
\end{array}
\label{Resistivity}
\end{equation}
The dissipative component of the conductivity tensor following from
this expression is given by:
\begin{equation}
{\rm Re}\sigma_{\rm xx} = \frac {n|e|c}B 
\frac{\gamma\omega_c^2\left(\omega^2+\omega_c^2+\omega_c^2\gamma^2\right)}
{\left(\omega^2-\omega_c^2-\omega_c^2\gamma^2\right)^2 +4\gamma^2\omega_c^2\omega^2},
\label{Sigma}
\end{equation}
where $\gamma = {\rm Re}\eta/m\omega_c$ is small in the superclean case
in accordance with Eq.~(\ref{Gamma}). At small frequencies 
Eq.~(\ref{Sigma}) can be seen to describe the cyclotron resonance. 
The cyclotron resonance in superconductors was observed experimentally in
Ref.~\onlinecite{Kirrai92} and was described by Ref.~\onlinecite{Hsu93} in the
framework of the kinetic equation in the $\tau$-approximation. 
Note that the cyclotron resonance
rendered by this expression is sharp if $\gamma \ll 1$ or exactly
in the conditions of superclean case.
Note also that as Eq.~(\ref{Sigma}) is a consequence of quite
general equation of motion of condensate (\ref{ParticularEq}).
In the derivation of Eq.~(\ref{Sigma}) we have neglected by the 
imaginary part of $\eta$ in comparison to $m\omega$. It is possible 
to do so because ${\rm Im}\eta$ has to vanish in the absence
of impurities. In the opposite case it would shift the position 
of the cyclotron resonance, violating the Kohn's theorem\cite{Kohn61},
as it was mentioned before.

Let us analize the influence of pinning on the motion of the
condensate. We will ignore the dissipation for a moment.
If pinning is present the pinning force has to be added to the total force
acting on the condensate in Eq.~(\ref{ParticularEq}):
\begin{equation}
\bbox{F}_p = -i\alpha \bbox{v}/\omega.
\end{equation}
Here $i\bbox{v}/\omega$ is the displacement of the vortex lattice 
(assuming it to be rigid), $\alpha$ is the pinning parameter, related
to the critical current $j_c$ by
\begin{equation}
\alpha = \frac {Bj_c}{cn_s\xi} \equiv m\omega_p^2.
\end{equation}  
In the last expression we have introduced the pinning frequency
$\omega_p$. This frequency can be related to the critical velocity 
$v_p \equiv j_c/n_s|e|$ and the cyclotron frequency by
\begin{equation}
\omega_p^2 = \frac {v_p}{\xi}\omega_c.
\end{equation}
Eqs.~(\ref{EquationWithVl}) and (\ref{ParticularEq}) give the following expression
for the modified by pinning frequency of the cyclotron resonance:
%
\begin{equation}
\tilde{\omega} = \frac{\omega_c^2+\omega_p^2}{\omega_c}
\label{CyclotronPinning}
\end{equation}
In the limit $\omega_p \rightarrow 0$, $\tilde{\omega} \rightarrow \omega_c$.
At a non-zero $\omega_p$ the cyclotron resonance occurs 
at a frequency larger then $\omega_c$ (Ref. \onlinecite{Hsu93}). 
At the same time the equation of vortex motion, Eq.~(\ref{VsMinusVl})
\begin{equation}
\bbox{v}_s - \bbox{v} = i \bbox{\hat{z}} \times \bbox{v} \omega_p^2/\omega \omega_c.
\label{VsMinusVlPin}
\end{equation}
Therefore the pinning is important if $\omega \ll \omega_p^2/\omega_c$.

Pinning can be ignored at large frequencies ($\omega \gg \tilde{\omega}$) or
if velocity is larger than $j_c/N_s|e|$. Then Eq.~(\ref{Sigma}) gives
\begin{equation}
{\rm Re} \sigma_{\rm xx} \approx {\rm Re} \rho_{\rm xx} 
\left(\frac {\omega_c}{\omega}\right)^2 
\left(\frac {n|e|c}{B}\right)^2 
\end{equation}
Using this expression one can relate the resonance structure
arising in ${\rm Re} \rho_{\rm xx}$ near $\omega \approx \omega_0$
to that in ${\rm Re} \sigma_{\rm xx}$. This resonance is described in the
next Section.

%
%
%
%

\section{Resonances at high frequency and small amplitude.}
\label{Resonances}

In this Section we will assume the displacement of the vortex due to the
driving field to be much smaller than the smallest scale at hand 
$\lambda_{\rm F}$. In this case to calculate absorption 
it is only natural to employ the Fermi golden rule.  
The absorption averaged over a large interval of $\omega$
in this linear response formalism is of the same order of magnitude as 
given by Eq.~(\ref{sigmaKK}).
However at the frequencies equal to multiples of $\omega_0$ 
resonances occur in the sample.
The aim of this Section is to calculate the absorption near these 
resonances.

Eq.~(\ref{GoldenRule}) emphasizes that at the conditions of resonance
the energies of two states $E_n$ and $E_m$ should be  close to the
multiples of $\omega_0$. This implies that they should not be disturbed
strongly from the values in the absence of impurity given by 
Eq.~(\ref{Energy0}). Therefore near resonances
the main contribution to the integral in Eq.~(\ref{many}) comes
from the regions far from the center of the vortex 
$\theta \xi \sim a_0 \ll \bar{a}$. In these regions one can
treat the influence of impurity perturbatively.
In the first order of the impurity potential we obtain the
following expressions for the correction to the 
energy of excitation from Eq.~(\ref{Energy})
\begin{equation}
\begin{array}{ll}
{\Delta E_n}&{ \equiv E_n-E_{0n} \approx (-1)^nI_1(\bar{a})} 
\end{array}
\end{equation}
At these conditions the transition probability in the first non-vanishing order in
the impurity potential can be determined  taking as wavefunctions the states of the
hamiltonial with no impurity [Eq.~(\ref{WaveFunctions})]. 
Eq.~(\ref{HamiltDeriv}) then gives
\begin{equation}
\left< m\right| \delta {\hat{\cal H}} \left| n \right> 
\approx 2 k_{\rm F} x_0\cos\phi 
\left\{
\begin{array}{ll}
{I_2(\bar{a}),}&{{\rm even}~n+m,} \\ \\
{I_1(\bar{a}),}&{{\rm odd}~n+m.}
\end{array} 
\right.
\label{MatrixElement}
\end{equation}
Here $\phi$ is the angle between $\bbox{\bar{a}}$ and $\bbox{x}_0$.
The golden rule expression (\ref{GoldenRule}) can be readily rewritten
as follows
\begin{equation}
\begin{array}{ll}
{Q}&{\displaystyle =n_{\rm i}n_{\rm v} \frac{\pi\omega}{2} 
\int d\phi \bar{a}d\bar{a} \sum_{n, m}\left[ f_n-f_m \right]} \\ \\
{}&{\displaystyle \left| 
\left< m\right| \delta {\hat{\cal H}} \left| n \right>
\right|^2
\delta \left[ \Delta \omega - I_1\left\{ (-1)^n - (-1)^m \right\}\right]},
\end{array}
\label{GeneralAbsorption}
\end{equation}
where $\Delta \omega = \omega - l\omega_0 < \omega_0/2$, with $l$ being integer, 
is the deviation of the frequency from the resonance, and index of the 
matrix element should be chosen in accordance with Eq.~(\ref{MatrixElement}).

Looking at the $\delta$-function in this expression one immediately sees
that the answer should be quite different for $\omega$ close to 
even and odd multiples of $\omega_0$ (for even and odd $l$).
Indeed if $l$ is even the resonant $\delta$-function reduces to 
$\delta(\Delta \omega )$. Therefore the condition of resonance is the same 
at {\em any} position of impurity relative to the center of vortex.
Hence $\delta(\Delta \omega )$ appears as a factor in front of the
expression for absorption. Therefore the resonances are
very {\em sharp} in this case.

In the opposite case, when $l$ is odd, the resonances can occur even
if $\Delta \omega$ is not zero, due to the shift of energy levels by
the impurity. Thus for odd $l$ the resonances are {\em broadened} by the
presence of impurity. This difference between even and odd $l$
is a consequence of the mentioned in Sec.~\ref{Excitations} property of the 
spectrum. The series of odd and even levels are shifted
with respect to each other by the impurity, preserving equal spacing of 
$2\omega_0$ within each series. Thus the resonances at even frequencies occur
within each series of levels and are sharp, while the perturbation at odd
frequencies mixes even and odd levels, making resonance broader.
Below we study these two cases separately.

{\em A) Odd $\omega/\omega_0$.}

Eq.~(\ref{GeneralAbsorption}) for this case can be rewritten as follows
\begin{equation}
\begin{array}{ll}
{Q}&{\displaystyle =n_{\rm i}n_{\rm v} \frac{2\pi^2\omega^2}{\omega_0}
k_{\rm F}^2 x_0^2 \int \bar{a}d\bar{a}
 \left( \frac{\Delta \omega}{2}\right)^2 } \\ \\
{}&{\displaystyle \times 
\delta \left[ \Delta \omega - 2\left|I(\bar{a})\right|
\sin \left(2k_{\rm F}\bar{a}\right)\right].}
\end{array}
\end{equation}
The resonances in the integrand of the last expression
occur very frequently (once in $\lambda_{\rm F}/4$). 
One can average contribution from these resonances over much larger
intervals to obtain an integral over a slowly varying integrand: 
\begin{equation}
\begin{array}{ll}
{Q}&{\displaystyle =n_{\rm i}n_{\rm v} \frac{\pi\omega^2}{2\omega_0}
k_{\rm F}^2 x_0^2 
\Delta \omega^2} \\ \\
{}&{\displaystyle \times \int \frac{\bar{a} d\bar{a}}
{\sqrt{\left[2\left|I\left(\bar{a}\right)\right|
\right]^2-\Delta \omega^2}} .}
\end{array}
\label{OddOmegaIntegral}
\end{equation}
The integral can be evaluated by changing the variable of integration
from $\bar{a}$ to $\left| I \right|$. They are related by Eq.~(\ref{Is}). 
The function $a\left(\left| I\right|\right)$ can be found
asymptotically in two limiting cases
\begin{equation}
a\left(\left| I\right|\right)=\left\{
\begin{array}{ll}
{\displaystyle \frac{2C^2\theta}{\pi k_{\rm F} |I| m} ,}
&{\displaystyle |I| \gg \theta \omega_0} \\ \\
{\displaystyle \frac{v_{\rm F}}{2\Delta_{\infty}} \ln 
\left( \frac{\theta \omega_0}{|I|} \right) ,}
&{\displaystyle |I| \ll \theta \omega_0}.
\end{array}
\right.
\end{equation}
Here
\begin{equation}
\theta \equiv m\tilde{V}\left( 2k_{\rm F}\right) \ll 1
\end{equation}
is the Born parameter of the impurity.
Calculating the absorption in these limiting cases one obtains
\begin{equation}
Q=n_{\rm i} n_{\rm v}v^2 \frac{2C^4\theta^2}{m^2 \omega_0 \Delta \omega},
\end{equation}
for $\theta \omega_0 \ll \Delta \omega \ll \omega_0$
and 
\begin{equation}
Q=n_{\rm i} n_{\rm v}v^2
\frac{\pi^2\varepsilon_{\rm F}\Delta\omega}
{2\Delta_{\infty}^2\omega_0}\ln\left( \frac {\theta\omega_0}{\Delta \omega}
\right),
\end{equation}
for $\Delta \omega \ll \theta \omega_0$.
It is convenient to express this answer in the units of absorption
obtained from the kinetic equation. 
Assume that $\tau$ is determined by scattering on short range impurities:
\begin{equation}
\tau^{-1}=2 \pi n_{\rm i} \nu \left| \tilde{V}\left( 2k_{\rm F}\right) \right|^2,
\end{equation}
with $\nu=m/\pi$ being 2D density of states in the layers
Then from Eq.~(\ref{sigmaKK}) one obtains
\begin{equation}
Q_{\rm K}=\frac{B}{2 n_s |e| c}\frac{j^2 }{\omega_0\tau}=
n_{\rm i} n_{\rm v}v^2 \theta^2  
\frac{\varepsilon_{\rm F}}{\omega_0}.
\label{KinEq}
\end{equation}
To simplify our consideration we will accept the Kramer-Pesch ansatz,
expressed by Eqs.~(\ref{KramersK}) and (\ref{KramersO}), after what we
obtain 
\begin{equation}
\frac{Q}{Q_{\rm K}} =
\displaystyle \frac{1}{\displaystyle 2\ln\left( \frac{T_c}{T}\right)} 
\frac{\omega_0}{\Delta \omega},
\label{ResonanceAnswer1}
\end{equation}
for $\theta \omega_0 \ll \Delta \omega \ll \omega_0$ and
\begin{equation}
\frac{Q}{Q_{\rm K}} =
\frac{\pi^2}{8}\ln\left( \frac{T_c}{T} \right)
\frac{\Delta \omega}{\theta^2\omega_0} 
\ln\left( \frac{\theta \omega_0}{\Delta \omega} \right),
\label{ResonanceAnswer2}
\end{equation}
for $\Delta \omega \ll \theta \omega_0$.

It can be noticed from these equations that 
when the frequency approaches the resonance
the absorption first increases in accordance with 
Eq.~(\ref{ResonanceAnswer1}) and then decreases
according to Eq.~(\ref{ResonanceAnswer2}).
The former expression therefore describes resonance near $\omega_0$,
while the latter describes the antiresonance. 
The numerical evaluation of Eq.~(\ref{OddOmegaIntegral}) shows that
the absorption reaches its maximum of ${\rm max} (0.2 / \theta; 0.6)$ 
at point $\Delta \omega \approx 0.5 \theta \omega_0$. 
If $\theta > 0.3$ the resonant part is not very well pronounced. On the other  
hand the antiresonace at $\Delta \omega \lesssim 0.5 \theta \omega_0$
exists independently of the value of the Born parameter.
An antiresonant behavior has been described earlier in Ref.~\onlinecite{Hsu93}.
The width of the antiresonance obtained in this reference is $1/\tau$.
We obtain for the width of the antiresonance $\Delta \omega \approx 0.5 \theta\omega_0$.
The discrepancy is the consequence of the mentioned failure of the 
$\tau$-approximation in (quasi)two dimensions.

Another perturbation causing the resonance could be due to the inertial forces.
This perturbation is important because it exists even in the 
cores with no impurities.
The corresponding correction to the hamiltonian, similarly to (\ref{HamiltDeriv}), is
\begin{equation}
\delta \hat{\cal H}_{\rm Inert}
=\left(
\begin{array}{ll}
{\displaystyle \bbox{\hat{p}} \bbox{v}_s; } & {\displaystyle \bbox{x}_0 \nabla \Delta}
 \\ \\
{\displaystyle \bbox{x}_0 \nabla \Delta^*; }&{\displaystyle \bbox{\hat{p}} \bbox{v}_s}
\end{array}
\right).
\label{HamiltInert}
\end{equation}
Here the diagonal terms represent the Doppler shift of the energy 
of the quaziparticles caused by the condensate moving with velocity $\bbox{v}_s$
(Ref. \onlinecite{Kopnin97}), while the off-diagonal terms are associated with 
the motion of vortex itself.\cite{Kopnin78} The matrix element 
of (\ref{HamiltInert}) calculated between the pure states relevant to 
the neighborhood of the resonance can be rewritten as follows:
\begin{equation}
\left< m\right| \delta \hat{\cal H}_{\rm Inert} \left| n \right> 
= \left< m\right| \bbox{\hat{p}}\left( \bbox{v}_s - \bbox{v}
\right) \sigma_0 \left| n \right>.
\end{equation}
Here $\sigma_0$ is the unit Pauli matrix.
In this equation we used the identity:
\begin{equation}
\left< m\right| \sigma_1 \bbox{x}_0 \nabla \Delta \left| n \right> =
\left< m\right| [\nabla; \hat{H}] \left| n \right> = 
\left< m\right|  \bbox{\hat{p}} \bbox{v} \sigma_0\left| n \right>.
\end{equation}
The absorption due to this perturbation is given by:
\begin{equation}
Q_{\rm Inert} = n_v n_s \frac{\omega_0}{4\tau_{\varepsilon}} 
\frac {\left( v - v_s \right)^2}{\Delta \omega^2 + \tau_{\varepsilon}^{-2}},
\end{equation}
where $v - v_s$ is given by either Eq.~(\ref{VsMinusVl}) or by 
Eq.~(\ref{VsMinusVlPin}). In case if pinning is negligible 
$\bbox{v}_s$ is very close to $\bbox{v}$, as it follows from Eq.~(\ref{VsMinusVl}).
Taking an estimate for $\eta$ from Eq.~(\ref{ResonanceAnswer2}) at 
$\Delta \omega \sim \tau_{\varepsilon}^{-1}$ the absorption due to the 
inertial forces can then be estimated as
\begin{equation}
Q_{\rm Inert}/Q_{\rm K} \sim 1/ \omega_0^2 \tau \tau_{\varepsilon}.
\end{equation}
This is just a small correction to Eq.~(\ref{ResonanceAnswer2}).
The significant reduction of the absorption is a consequence of the Galellian 
invariance.



%
%

{\em B) Even $\omega/\omega_0$.}

In this case due to homogeneous broadening of the levels the 
$\delta$ function in Eq.~(\ref{GeneralAbsorption}) is replaced by
the following expression:
\begin{equation}
\delta(\Delta \omega) = \frac {1}{\displaystyle \pi \tau_{\varepsilon}
\left(\Delta \omega^2 + \tau_{\varepsilon}^{-2}\right)}.
\end{equation}
Eq.~(\ref{Is}) gives the following result for the absorption
\begin{equation}
Q=n_{\rm i}n_{\rm v}\frac{4C^4\theta^2}{m^2\omega_0}\delta(\Delta \omega)
\ln\left( \frac{1}{\theta}\right).
\label{even_answer}
\end{equation}
Using the Kramer-Pesch ansatz (\ref{KramersK}) and (\ref{KramersO})
one readily obtains 
\begin{equation}
\frac{Q}{Q_{\rm K}}=\displaystyle 
\frac{\displaystyle \ln\left( \frac{1}{\theta}\right)}
{\displaystyle \ln\left( \frac{T_c}{T}\right) }
\delta(\Delta \omega)
\label{EvenAnswerKramers}
\end{equation}
Note, that the absorption averaged over frequency for $\omega > \omega_0$
is the same in both even and odd cases and equals to
\begin{equation}
\frac{\bar{Q}}{Q_{\rm K}}=\frac 1{Q_{\rm K}\omega_0}
\int_{(l-1/2)\omega_0}^{(l+1/2)\omega_0}d\omega Q(\omega)=
\frac{\displaystyle \ln\left( \frac{1}{\theta}\right)}
{\displaystyle \ln\left( \frac{T_c}{T}\right) }
\end{equation}

%
%
%
%

\section{Small amplitude low frequency absorption.}
\label{Small_amplitude}

At a frequency much smaller than $\omega_0$ absorption is only
possible if the impurity makes two levels approach each other
to a distance smaller than $\omega$. This is possible if
impurity is situated within dissipative region described 
in Sec.~(\ref{Excitations}).
From Eq.~(\ref{AnticrossingEnergy}) it follows that the condition 
$\delta E < \omega$,
$\omega \rightarrow 0$ can be satisfied in a narrow region 
adjacent to $\bar{a} = a_0$. In this region odd and even levels 
periodically anticross, approaching each other to  small
distances.

When two levels are anomalously close all others can be ignored
and the two-state system can be described by a hamiltonian
\begin{equation}
H = \left( 
\begin{array}{cc}
{\protect{\alpha }}&{\protect{\delta}} \\ \\
{\protect{\delta}}&{-\protect{\alpha}}
\end{array}
\right),
\label{2LevHamiltonian}
\end{equation}
where $\delta > 0$ is a slowly varying function of the $\bar{a}$, while $\alpha$
changes rapidly and $\alpha=0$ at the point of level anticrossing. 
The eigenvalues of this two-level system are given by
\begin{equation}
\varepsilon = \pm \sqrt{\delta^2 + \alpha^2}.
\label{2LevEnergy}
\end{equation}
Comparing this equation with Eq.~(\ref{Energy}) in the neighborhood of the
anticrossing one obtains the following expression for $\alpha$ and $\delta$
\begin{equation}
\begin{array}{l}
{\displaystyle \delta = \frac{\delta E}{2}} \\ \\
{\displaystyle \alpha \approx I_2 \approx \frac{2k_{\rm F} \delta a \omega_0} {\pi} },
\end{array}
\end{equation}
where $\delta E$ is given by Eq. (\ref{AnticrossingEnergy}) 
and $\delta a$ is the distance from the point of anticrossing.
The eigenmodes of the hamiltonian are described by
\begin{equation}
a_{\pm} = \left( 
\begin{array}{l}
{\displaystyle 
\sqrt{\frac{|\varepsilon|\pm \delta}{2|\varepsilon|}}
} \\ \\
{\displaystyle 
\pm \sqrt{\frac{|\varepsilon|\mp \delta}{2|\varepsilon|}}
}.
\end{array}
\right)
\end{equation}
The upper and lower signs in this expression pertain to the states 
with positive and negative energy in Eq.~(\ref{2LevEnergy}).
Now we can apply the
golden rule expression for absorption (\ref{many}) to this
two-level system:
\begin{equation}
\begin{array}{ll}
{Q}&{\displaystyle = 
n_{\rm i}n_{\rm v}\omega \int d \phi \bar{a} d \bar{a} \frac{\pi}{2}
x_0^2\cos^2\phi 
} \\ \\
{}&{\displaystyle \times 
\left| \left< +\right| \frac{\partial H}{\partial a} \left| -\right>\right|^2 
\delta\left[\omega-2\varepsilon
\left(\bar{a}\right) \right]},
\end{array}
\label{StartingForLowFreq}
\end{equation}
with
\begin{equation}
\left< +\right| \frac{\partial H}{\partial a} \left| -\right>=
\frac 2{\pi} k_{\rm F}\omega_0 \frac {\delta}{|\varepsilon|}.
\label{2LevMatrix}
\end{equation}
One can average over frequently occurring resonances in the integrand
of Eq.~(\ref{StartingForLowFreq}) to obtain an integral over a slowly
varying function
\begin{equation}
\begin{array}{ll}
{Q}&{\displaystyle = 
2 n_{\rm i}n_{\rm v} x_0^2k_{\rm F}^2 \omega_0 a_0 
} \\ \\
{}&{\displaystyle \times 
\int d \bar{a} \frac{\delta^2(\bar{a})}
{\sqrt{\displaystyle\frac{\omega^2}{4} - \delta^2(\bar{a})}}
}.
\end{array}
\end{equation}
Changing the variables of integration from $\bar{a}$ to $\delta$ using 
Eq.~\ref{AnticrossingEnergy}, i.e. using 
\begin{equation}
d\bar{a} = 4d\delta d\bar{a}/d|I| = 2\pi a_0 d\delta / \omega_0,
\label{DOS}
\end{equation}
we eventually obtain
\begin{equation}
\frac{Q}{Q_{\rm K}}=\frac{\pi^2}{\displaystyle 16\ln\left(\frac{T_c}{T}\right)}.
\label{AnswerAtSmallFreq}
\end{equation}
To obtain this answer we have employed the Kramer-Pesch ansatz
(\ref{KramersK}) and (\ref{KramersO}).

\begin{figure}
\centerline{
\psfig{file=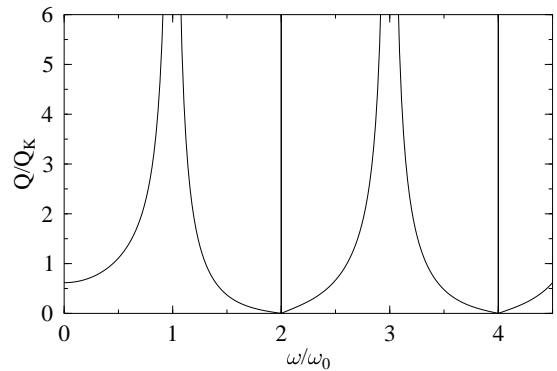,height=1.9in,bbllx=60pt,bblly=103pt,bburx=489pt,bbury=369pt}
}
\vspace{0.1in} 
\setlength{\columnwidth}{3.2in}
\centerline{\caption{Absorption as a function of frequency determined by 
\protect{Eq.~(\ref{GeneralFormula})}. 
At add frequencies only resonant behavior is shown.
The sharp resonances at even frequencies 
[\protect{Eq.~(\ref{EvenAnswerKramers})}] are shown by vertical lines. 
The non-zero limit\protect{$Q(\omega \rightarrow 0)$} is evaluated in this Section.
\label{fig30}
}}
\vspace{-0.1in}
\end{figure}

Eq.~(\ref{many}) can also be evaluated for general $\omega$. The result reads
\begin{equation}
\begin{array}{ll}
{\displaystyle \frac{Q}{Q_{\rm K}}}
&{ = \displaystyle \frac{\pi \omega_0}{\displaystyle 2 
\ln\left( \frac{T_c}{T} \right) 
\omega}
\left[
2{\rm floor} \left( \frac{\omega}{2\omega_0} \right)+1
\right]
} \\ \\
{}&{\displaystyle \times 
\left|\frac{\displaystyle \cos^3\left( \frac{\pi\omega}{2\omega_0} \right)
-\cos \left( \frac{\pi\omega}{\omega_0} \right)
}
{\displaystyle \sin\left( \frac{\pi\omega}{\omega_0} \right)}\right|},
\end{array}
\label{GeneralFormula}
\end{equation}
where ${\rm floor}(x)$ is the maximum integer smaller or equal than $x$.
Note that both asymptotics
given by Eqs.~(\ref{ResonanceAnswer1}) and (\ref{AnswerAtSmallFreq})
can be obtained from this expression in the limits
$\omega \rightarrow 0$ and $\omega \rightarrow (2n+1)\omega_0,~n=0,1,\ldots$
respectively.  
This result is shown in Fig.~\ref{fig30}. 

%
%

As it was mentioned in the previous Section, due to numerical reasons,
the absorption at odd frequencies ($\omega_0, 3\omega_0, \ldots$) 
does not raise above 
${\rm max} (0.2 / \theta; 0.6)$. The resonances at odd frequencies
can therefore be pronounced only if $\theta \lesssim 0.3$. 
We expect that in practice $\theta \lesssim 1$.
Eq.~(\ref{GeneralFormula})
and Fig.~\ref{fig30} can therefore be understood only asymptotically 
in the limit $\theta \ll 1$.

%
%
%
%

\section{Non-linear effects in the absorption}
\label{Non-linear}

The non-linear effects in the absorption are interesting since in many cases
they depend on the energy relaxation time of excitations. 
Therefore they can be used for studying this quantity. 
We attribute the non-linearities to two different phenomena. 
The first phenomenon is
the saturation of absorption due to the redistribution of level occupations. 
It corresponds to the deviations from the golden rule expression (\ref{one}).
This leads to a {\em decrease} of the dissipative component of resistivity
with respect to a linear response result.
The second group of non-linear phenomena 
includes the effects associated with direct transitions
between levels due to their anticrossings. They {\em increase} the 
absorption.

Consider the deviations from the golden rule expression first. In a two-level
system the resonance with external field brings about the rotation of the population
of the levels with Rabi frequency $\Omega_{\rm R}$.
For the cases described in Sec.~\ref{Resonances} 
and \ref{Small_amplitude} the Rabi frequency is given by:
\begin{equation}
\begin{array}{ll}
{\Omega_{\rm R}}&{=\displaystyle 
\left| \left< m\right| \delta {\cal{H}} \left| n \right> \right|} \\ \\
{}&{\displaystyle \sim \omega_0 \frac{x_0}{\lambda_{\rm F}} }.
\end{array}
\label{OmegaR}
\end{equation}
If the perturbation is applied during a time interval longer than $1/\Omega_{\rm R}$,
the absorption saturates. In this case the system can absorb only if
there are inelastic relaxation mechanisms present. If the corresponding relaxation 
time is $\tau_{\varepsilon}$, the energy $\omega$, corresponding to the 
external field frequency, is absorbed once in $\tau_{\varepsilon}$.
This is in contrast to the golden rule expression, from which
this time is seen to be $1/\Omega_{\rm R}$. To account for this this effect 
the absorption should be renormalized as follows
\begin{equation}
Q \rightarrow \frac{Q}{\sqrt{1+\Omega_{\rm R}^2\tau_{\varepsilon}^2}}.
\end{equation}
The denominator in this expression is large if 
$v/v_c \gg \omega/\omega_0^2\tau_{\varepsilon}$. The absorption
in this case is
\begin{equation}
Q \sim Q_{\rm K} \frac{v_c\omega}{v\omega_0^2\tau_{\varepsilon}}.
\label{RegionD}
\end{equation}

\begin{figure}
\centerline{
\psfig{file=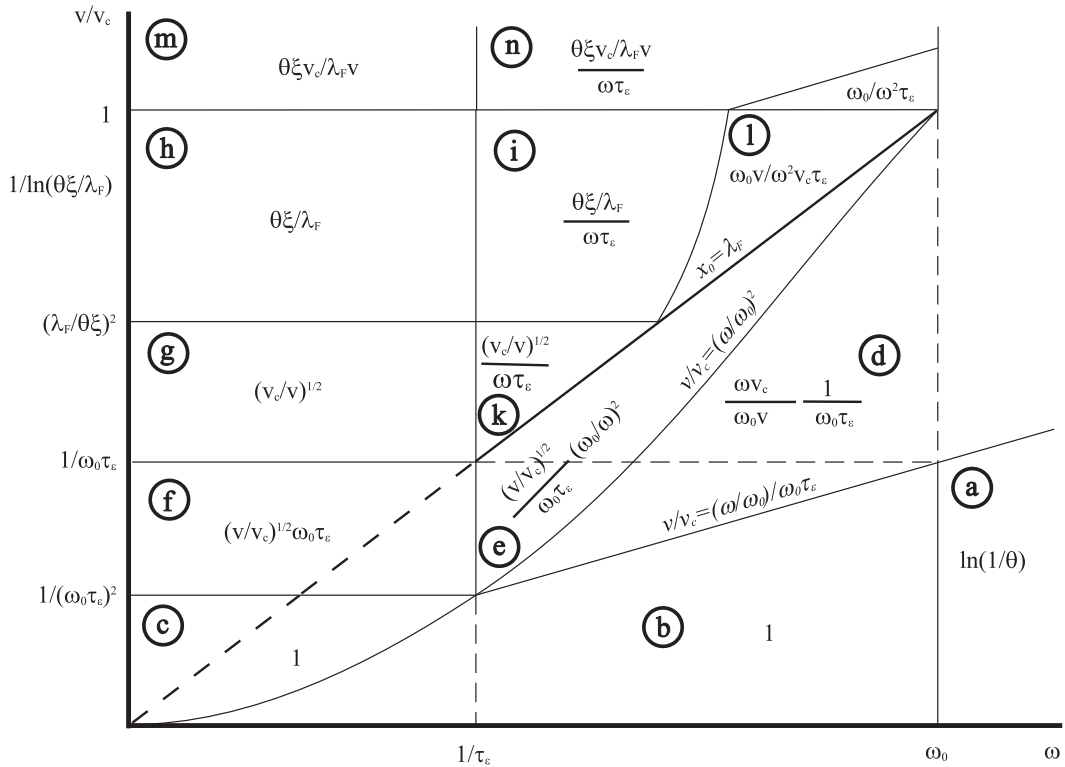,height=4in}}
\vspace{0.1in} 
\setlength{\columnwidth}{3.2in}
\centerline{\caption{The diagram of absorption.
The values of \protect{$Q/Q_{\rm K}$} are mapped for different
regions in (\protect{$v/v_c$, $\omega$}) plane. The equations for the boundaries
between these region are made italic for convenience.
\label{fig40}
}}
\vspace{-0.1in}
\end{figure}

The non-linear effects are relatively abundant. To visualize them we
draw a diagram in the space of parameters $v$ and $\omega$. 
This diagram is shown in Fig.~\ref{fig40}. It displays the absorption
measured in the units of $Q_{\rm K}$ given by Eq.~(\ref{KinEq}). 

The golden rule absorption considered in the
previous two Sections is situated in sectors (a) and (b)
of Fig.~\ref{fig40}. The saturation
of the golden rule expression becomes important on the line 
$v/v_c = \omega/\omega_0^2\tau_{\varepsilon}$. Eq.~(\ref{RegionD}) 
determines the absorption in sector (d).

\end{multicols}

%
%

\begin{multicols}{2}

We now turn to the discussion of the influence of the Zener transitions
between anticrossing levels. These transitions occur within the dissipative region,
where some levels can be anomalously close. 
In this case the system of two almost degenerate states 
can be described by a two-level hamiltonian (\ref{2LevHamiltonian}), and their energies
are given by Eq.~(\ref{2LevEnergy}). The occupations of such a two-level system 
before the anticrossing $(p_1, p_2)$ and after the anticrossing $(p_1', p_2')$
are related as follows:\cite{LL3}
\begin{equation}
\left( 
\begin{array}{l}
{p_1'} \\ {p_2'}
\end{array}
\right)
=
\left( 
\begin{array}{ll}
{1-W;}&{W} \\ {W;}&{1-W}
\end{array}
\right)
\left( 
\begin{array}{l}
{p_1} \\ {p_2}
\end{array}
\right).
\end{equation}
The transition probability to the other level relevant for the absorption is
\begin{equation}
W=\exp\left( -\frac {\pi^2 \delta E^2} 
{\displaystyle 4 k_{\rm F}\omega_0 v} \right).
\label{ZenerProbability}
\end{equation}
Here $\delta E$ is the smallest distance between levels 
determined by Eq.~(\ref{AnticrossingEnergy}).

Zener transitions can occur if $W \sim 1$. This determines the width
of the dissipative region as well as the number of 
effectively working anticrossings there $N$.
Indeed, if an impurity is situated in a ring of width $\lambda_{\rm F}N$
and radius $a_0 \sim \theta \xi$ it can produce Zener transitions.
We obtain from Eq.~(\ref{AnticrossingEnergy})
\begin{equation}
\delta E \sim \frac{d|I|}{da} \lambda_{\rm F} N \sim 
\frac {\omega_0 } {\theta \xi} \lambda_{\rm F} N. 
\label{deltaEApprox}
\end{equation}
Then from Eq.~(\ref{ZenerProbability}) we have
\begin{equation}
N \sim \frac{\theta \xi}{\lambda_{\rm F}}\sqrt{\frac {v} {v_c} }.
\label{NNN}
\end{equation}
One can distinguish two cases, depending on whether $N$ is large or small.
If $N$ is large, an impurity passing through the dissipative region
brings about many transitions. The opposite case $N \ll 1$ can
be realized if velocity is small $v\ll v_c (\lambda_{\rm F}/\theta\xi)^2$.
Below we first consider this latter case in some detail.

If $N \ll 1$, for a typical impurity 
Zener transitions are  improbable.
However if the Born parameter of the impurities has a little dispersion
$\delta \theta = \sqrt{\left<\theta^2\right>-\left<\theta\right>^2} \neq 0$,
the Zener transitions still can happen, yet on a very small fraction of 
impurities. Indeed, the current position of the level crossing $a$ 
is quantized in the units of $\lambda_{\rm F}/4$. 
The uncertainty of $a_0\sim\theta \xi$ is $\delta a_0 \sim \delta \theta \xi$. 
Let us look at Eq.~(\ref{AnticrossingEnergy}). 
If $\delta a_0 > \lambda_{\rm F}/4$, $\delta E$ in this 
expression can accidentally become zero.
This condition can also be rewritten as 
$\delta\theta > \lambda_{\rm F}/\xi \ll 1$. 
This assumption can therefore be realized even if the uncertainty of the Born
parameter is  small.

This can be reformulated in terms of an assumption about 
the distribution function of $\delta E$. In the case when 
$\delta \theta > \lambda_{\rm F}/\xi$ this distribution
function $p(\delta E)$ has a non-zero limit when $\delta E$ goes to zero.
In particular, taking Eq.~(\ref{deltaEApprox}) into account, we obtain
\begin{equation}
p(\delta E) = \frac {dN}{d[\delta E]} \sim \frac {\theta \xi }
{\lambda_{\rm F}\omega_0}.
\label{deltaEDOS}
\end{equation}
The expression for the total absorption Eq.~(\ref{many}) can be
rewritten in this case as follows ($\bar{a}\sim \theta\xi$, 
$d\bar{a}\sim x_0$)
\begin{equation}
Q \sim n_{\rm i}n_{\rm v} \theta \xi x_0 \int_0^{\infty} Q_1(\delta E) p(\delta E)
d[\delta E],
\label{manyZener}
\end{equation}
where $Q_1(\delta E)$ is the absorption due to one impurity driving
Zener transition on a couple of levels with the shortest separation $\delta E$.
Our task from now on will be to estimate $Q_1$.

For the following discussion it is convenient to consider two cases
$x_0 \ll \lambda_{\rm F}$ and $x_0 \gg \lambda_{\rm F}$ separately.
In the first case an impurity can cause only one Zener transition per 
period, while in the second case there can be many.

{\em A) $x_0 \ll \lambda_{\rm F}$.}

Consider an impurity at a distance less than $x_0$ from one of the anticrossings.
Such an impurity can cause Zener transitions once in $\omega^{-1}$.
The absorbed energy, however, depends on the relation between the
characteristic time of Zener transition $\Delta t$ and the inelastic
relaxation time $\tau_{\varepsilon}$. The former can be estimated as
\begin{equation}
\Delta t \sim \frac {\delta E} {\omega_0}\frac{\lambda_{\rm F}}{v}
\end{equation} 
and is the time during which two levels are at a distance 
of the order of $\delta E$. The absorption rate due to one such impurity
can be estimated in three limiting cases as follows
\begin{equation}
Q_1 \sim
\left\{
\begin{array}{ll}
{\delta E W \omega, }     &     {\tau_{\varepsilon} \ll \Delta t} \\ \\ 
{\displaystyle \frac{v\tau_{\varepsilon}}{\lambda_{\rm F}}\omega_0 W \omega,}&
{ \Delta t \ll \tau_{\varepsilon} \ll \omega^{-1}} \\ \\
{\displaystyle \frac {x_0}{\lambda_{\rm F}}\omega_0 W \frac 1{\tau_{\varepsilon}},}&
{\Delta t \ll \omega^{-1} \ll  \tau_{\varepsilon}}
\end{array}
\right.
\label{oneZener}
\end{equation}
As it has been mentioned above 
this is true if the impurity is at a distance smaller than $x_0$ from
the anticrossing. In the opposite case the Zener transitions cannot
occur and the absorption is negligible. 
Eqs.~(\ref{manyZener}) and (\ref{oneZener}) then result in three limiting cases
\begin{equation}
Q \sim Q_{\rm K}
\left\{
\begin{array}{ll}
{\displaystyle \sqrt{\frac{v}{v_c} } \omega_0 \tau_{\varepsilon}; }     &     
{\displaystyle \omega \tau_{\varepsilon} \ll 1,~ 
\frac {v}{v_c}\gg \frac{1}{\omega_0^2\tau_{\varepsilon}^2} } \\ \\ 
{ 1; }     &     
{\displaystyle \omega \tau_{\varepsilon} \ll 1,~ 
\frac {v}{v_c}\ll \frac{1}{\omega_0^2\tau_{\varepsilon}^2} } \\ \\ 
{\displaystyle  \frac{\omega_0^2}{\omega^2} \frac{1}{\omega_0\tau_{\varepsilon}} 
\sqrt{\frac{v}{v_c}}; }&{\displaystyle  \omega\tau_{\varepsilon} \gg 1.}
\end{array}
\right.
\end{equation}
These limiting cases correspond to the regions labeled (f), (c), and (e) 
($\lambda_{\rm F} \gg x_0$) correspondingly in the diagram in Fig.~\ref{fig40}.

Note that for this case the answer does not change when velocity 
passes across the value $v=v_c\lambda_{\rm F}^2/\theta^2\xi^2$ where $N\sim 1$.
This is due to the fact that if $x_0 \ll \lambda_{\rm F}$ even for the 
case $N\gg 1$ the moving impurity cannot produce more than one Zener
transition per period $2\pi/\omega$. This happens not to be the case
for $x_0 \gg \lambda_{\rm F}$, when the boundary 
$v=v_c\lambda_{\rm F}^2/\theta^2\xi^2$ does exist.

{\em B) $x_0 \gg \lambda_{\rm F}$.}

As it has been mentioned above for this case the division into
small and large velocities is essential. 
We start from small velocities $v \ll v_c\lambda_{\rm F}^2/\theta^2\xi^2$,
as in the experimental conditions\cite{Matsuda94} this is the region
where the non-ohmic absorption can first be observed.

For small velocities a consideration analogous to 
Eq.~(\ref{oneZener}) leads to
\begin{equation}
Q_1 \sim
\frac 1{\omega \tau_{\varepsilon}} \tanh \left(\omega \tau_{\varepsilon}
\right) \left\{
\begin{array}{ll}
{\delta E W \omega, }     &     {\tau_{\varepsilon} \ll \Delta t} \\ \\ 
{\displaystyle \frac{v\tau_{\varepsilon}}{\lambda_{\rm F}}\omega_0 W \omega,}&
{ \Delta t \ll \tau_{\varepsilon} \ll \lambda_{\rm F}/v } \\ \\
{\displaystyle \frac {x_0}{\lambda_{\rm F}}\omega_0 W
\frac 1{\tau_{\varepsilon}},}&
{\Delta t \ll \lambda_{\rm F}/v \ll  \tau_{\varepsilon}}
\end{array}
\right.
\label{oneZenerLrgAmp}
\end{equation}
The factor 
$\tanh \left(\omega \tau_{\varepsilon}\right)/\omega \tau_{\varepsilon}$
accounts for the saturation of absorption when 
$\omega \gg \tau_{\varepsilon}^{-1}$. This is analogous to the saturation
discussed in the Fermi golden rule case.
This results in 
\begin{equation}
Q \sim 
\frac {Q_{\rm K}}{\omega \tau_{\varepsilon}} \tanh \left(\omega \tau_{\varepsilon}
\right)
\left\{
\begin{array}{ll}
{\displaystyle \sqrt{\frac{v_c}{v} }, }     &     
{\displaystyle 1/\omega_0 \tau_{\varepsilon} \ll v } \\ \\ 
{\displaystyle \sqrt{\frac{v}{v_c}}, }     &     
{\displaystyle 1 / \omega_0^2 \tau_{\varepsilon}^2 
v \ll 1 / \omega_0 \tau_{\varepsilon}  } \\ \\ 
{\displaystyle  1, }&{v \ll 1 / \omega_0^2 \tau_{\varepsilon}^2.}
\end{array}
\right.
\end{equation}
Note that for small frequencies $\omega \ll 1/\tau_{\varepsilon}$ we
have reproduced two of the results from the previous case 
$x_0 \ll \lambda_{\rm F}$, namely, sectors (f) and (c) of the diagram
in Fig.~\ref{fig40}. For this reason the line $x_0 = \lambda_{\rm F}$
in this Figure is made dashed, as it does not divide physically different 
regions. We have also obtained a new result for sector (g). 

Let us now discuss the case $N \gg 1$. If the amplitude of vortex motion
$x_0$ exceeds $\lambda_{\rm F} N$, 
the impurity can cause $N$ Zener transitions per period of motion $2\pi/\omega$.
This implies that the excitations can 
{\em ballistically} propagate up in energy to the hight $\omega_0N$ above
the Fermi level. Therefore the power absorbed in one vortex 
can be written as follows
\begin{equation}
Q_1 \sim \frac 1 {\tau_{\varepsilon}} \tanh \left(\omega \tau_{\varepsilon}
\right) \omega_0 N^2.
\end{equation}
Hence, for the case $\lambda_{\rm F} N \ll x_0$ Eq.~(\ref{many})
results in ($\bar{a}\sim \theta \xi$, $d\bar{a}\sim x_0$)
\begin{equation}
\begin{array}{ll}
{Q}&{ \displaystyle \sim n_{\rm i}n_{\rm v} \theta \xi x_0 Q_1} \\ \\
{}&{\displaystyle \sim Q_{\rm K}
\left\{
\begin{array}{ll}
{\displaystyle \frac{\theta \xi}{\lambda_{\rm F}}, }&
{\displaystyle \omega \tau_{\varepsilon} \ll 1} \\ \\
{\displaystyle \frac{\theta \xi}{\lambda_{\rm F}\omega \tau_{\varepsilon}}, }&
{\displaystyle \omega \tau_{\varepsilon} \gg 1.}
\end{array}
\right.
}
\end{array}
\end{equation}
These answers correspond to the sectors (h) and (i) in Fig.~\ref{fig40}.

The equality $x_0 = \lambda_{\rm F} N$ takes place on the line 
$v/v_c=(\omega \theta \xi / \omega_0 \lambda_{\rm F})^2$. This line
is the boundary between sectors (i) and (l) of the diagram. 
In sector (l) $x_0 \ll \lambda_{\rm F} N$, and the maximum energy
above the Fermi level that excitations can reach is $\omega_0 x_0/\lambda_{\rm F}$.
Therefore 
\begin{equation}
Q_1 \sim \frac 1 {\tau_{\varepsilon}} \tanh \left(\omega \tau_{\varepsilon}
\right) \omega_0 \left( \frac{x_0}{\lambda_{\rm F}} \right)^2.
\end{equation}
Eq.~(\ref{many}) results in ($\bar{a}\sim \theta \xi$, 
$d\bar{a} \sim \lambda_{\rm F} N$)
\begin{equation}
\begin{array}{ll}
{Q}&{ \displaystyle \sim n_{\rm i}n_{\rm v} \theta \xi \lambda_{\rm F} N Q_1} \\ \\
{}&{\displaystyle \sim Q_{\rm K}
\left\{
\begin{array}{ll}
{\displaystyle \sqrt{v/v_c}, }&
{\displaystyle \omega \tau_{\varepsilon} \ll 1} \\ \\
{\displaystyle \sqrt{v/v_c} / \omega \tau_{\varepsilon}, }&
{\displaystyle \omega \tau_{\varepsilon} \gg 1.}
\end{array}
\right.
}
\end{array}
\end{equation}
These answers correspond to the sector (l) of Fig.~\ref{fig40}.

Finally we would like to discuss the crossover between the 
regions (h), (i), and (l) and the non-ohmic regimes at high velocities 
in the sectors (m) and (n).
When velocity exceeds the critical velocity $v_c$ the number of anticrossing
is restricted by the size of the dissipative region. 
It is therefore
\begin{equation}
N\sim \frac{\theta \xi}{\lambda_{\rm F}}
\end{equation}  
Therefore the answers in sectors (m) and (n) can be obtained by 
renormalization those in sectors (h) and (i) by the factor $v_c / v$.

In the case of small frequencies $\omega\tau_{\varepsilon} \ll 1$
one can give the following interpolation formula for the absorption
\begin{equation}
\begin{array}{ll}
{\displaystyle \frac{Q}{Q_{\rm K}} }&{\sim 1+\frac{\displaystyle 1}{\displaystyle
\sqrt{v_c/v}/\omega_0\tau_{\varepsilon}
+ \sqrt{v/v_c}   } } \\ \\
{} & {\displaystyle \frac{1}{\displaystyle \sqrt{v_c/v}/\omega_0\tau_{\varepsilon}
+ \theta \xi /\lambda_{\rm F}   } \frac 1{1+v/v_c} } 
\end{array}
\end{equation}
The second term arises due to the Landau-Zener transitions on rare impurities,
causing very small $\delta E$. Such impurities can therefore cause 
no more than one transition per period.
The third term is associated with the cascade
of Landau-Zener transitions. If 
$\omega_0 \tau_{\varepsilon} < \theta \xi/\lambda_{\rm F}$, then the 
third term is larger than the second and sectors (g) and (k) in Fig.~\ref{fig40}
do not exist.

%
%
%
%

\section{Summary and Conclusions}
\label{Conclusions}

In this work we have studied the influence of the discreteness of excitation levels
on microwave absorption in superclean layered superconductors. 
At  low amplitudes and low frequencies ($\omega \ll \omega_0$)
the absorption coincides with the result of Ref.~{\onlinecite{KK}}.
With increasing amplitude of vortex motion two non-linear effects are observed.
In the case $\omega \tau_{\varepsilon}\gg 1$ an increase of the amplitude
decreases the dissipative component of resistivity. 
This is due to the deviation of the occupation of the excitation levels 
from the Fermi distribution.
In the opposite case $\omega \tau_{\varepsilon}\ll 1$
an increase of the amplitude brings about an increase of the dissipative
component of resistivity due to Zener transitions.
Therefore this effect can be used to measure the inelastic relaxation time.

In Sec.~\ref{Motion} we discuss the conditions at which the cyclotron resonance
can be observed in superconductors. We find that the broadening 
of the resonant curve is $\omega_c/\omega_0\tau$. We conclude therefore
that the cyclotron resonance is sharp if the condition of superclean
case ($\omega_0\tau\gg 1$) is satisfied. This provides and independent way to 
measure this quantity. We also calculate the correction to the cyclotron frequency 
brought about by pinning.

For small amplitudes and large frequencies 
we obtain the series of antiresonances at the odd frequencies 
commensurate with the excitation level spacing [$(2n+1)\omega_0$, $n=0,~1,~2,~\ldots$],
and the series of resonances at even frequencies commensurate with the 
level spacing [$2n\omega_0$]. 
Therefore the resonant behavior of the vortex cores in superclean 
superconductors reveals an even - odd anomaly.
The existence of antiresonances at odd frequencies
is associated with the retraction of the effectively working near the resonance 
impurity from the vortex to the distances $a \gg \xi$. There it creates very small
matrix element of transition between vortex states.
On the other hand at even frequencies the effectively working near 
the resonance impurity resides inside the vortex core ($a \lesssim \xi$).

Let us discuss the assumptions that we have made. 
We assumed the temperature to be sufficiently small.
First, it has to be much smaller than $T_c$, so that
the relevant excitations are given by Caroli - de Gennes - Matrison theory
\cite{deGennes89}. Second, $\tau_{\varepsilon}(T)$ should be large enough
to satisfy the condition of the superclean case 
$\omega_{0} \tau_{\varepsilon}\gg 1$. 

We also adopted the model of disorder consisting of strong short-range
impurities with the Born parameter satisfying the condition
$\theta \gg (\omega_0\tau)^{-1/2}$. At this condition
there is no more than one impurity per vortex core per layer.
In the opposite case of white noise disorder potential the excitation levels can
be shown to be broadened by $\sqrt{\omega_0/\tau}$.
This is analogous to the case of Landau level broadening by the white
noise disorder considered in Ref.~\onlinecite{Ando}. In this
case we expect the absorption due to impurities at  low frequencies
($\omega \ll \omega_0$) to be exponentially small.~\cite{Ando_cyclotron,Fogler98}

We have considered 2D case assuming that tunneling between layers is  
small. In the presence of such tunneling the excitation levels with
no impurities are broadened into band. The width of the band is of the order
of $\delta \omega_0 \sim \omega_0 \epsilon$, 
where $\epsilon \ll 1$ is the anisotropy parameter. 
However, in the presence of impurity one level can leave the band
and become discrete. 
The virtual transition to the other layers produce only
small correction to the energy of this level and to our results.
This correction can be accounted for
by the renormalization of the impurity potential. 
However the antiresonances at $(2n+1)\omega_0$ and resonances at $2n\omega_0$
should be smeared in the presence of anisotropy. The finite homogeneous broadening
of levels $\tau^{-1}_{\varepsilon}$ should also add some 
broadening to the (anti)resonances. We therefore conclude that the (anti)resonances
should be broadened by ${\rm min}(\delta\omega_0,~\tau^{-1}_{\varepsilon})$.

We disregarded pinning of the vortex lattice by the impurities in
our consideration. It is possible to do so if the pinning does not
affect motion of the lattice. This happens 
when $\omega_p^2 < \omega\omega_c$, where $\omega_p$ is the
characteristic pinning frequency introduced in 
Section~\ref{Motion} [Eq.~(\ref{CyclotronPinning})]. 
Pinning can also be neglected if the current significantly exceeds the critical current.

We have assumed the s-wave pairing mechanism in our treatment.
In practice the superclean case can be realized in high-$T_c$
and organic superconductors. It is accepted to think that
these materials have a d-wave coupling mechanism. Therefore
it is important to study the absorption in d-wave superclean superconductors.
This study has been done by Kopnin and Volovik \cite{Kopnin97} 
in the framework of kinetic equation in the $\tau$-approximation. 
We expect however that the
differences between the results of kinetic equation and the microscopic
approach studied in this work persist in the d-wave superconductors.
Therefore we assume that some qualitative results obtained in  
our paper are applicable to these materials also. This 
matter requires a further study.

One of the known to us experiments on superclean samples is described in
Refs.~\onlinecite{Matsuda94}. 
This paper reports a large $\omega_0 \tau \approx 14$ observed in 
90K single-crystal YBCO sample. This quantity weakly depends on temperature
below 17K. Therefore we expect that at the conditions of weak temperature
dependence scattering on impurities provides the main mechanism 
of absorption.

As it follows from our results the study of the frequency 
dependence and/or non-linear effects of absorption it is
possible to determine the inelastic scattering time $\tau_{\varepsilon}$
even if it is much larger then the elastic one. 
Knowledge of $\tau_{\varepsilon}$ is  important for understanding
of the peculiarities of high-$T_c$ superconductors.\cite{Anderson91,Ioffe98}

%
%
%
%

\section{Acknowledgements}
\label{Acknowledgements}

The authors would like to express gratitude to Yu.~M.~Galperin, 
V.~B.~Geshkenbein, and B.~I.~Shklovskii for helpful discussions.
A.~K. was supported by NSF grant DMR-9616880, A.~L. by NSF grant DMR-9812340.

\appendix

%
%
%
%


\vspace{-0.2in}

\end{multicols}
\end{document}